\documentclass[10pt]{article}
\begin{document}
%%%%%%%%%%%%%%%%%%%%
\title{{\bf Hawking radiation from black holes in de Sitter spaces 
via covariant anomalies}}
%%%%%%%%%%%%%%%%%%%%
\author{
{Sunandan Gangopadhyay$^{}$\thanks{sunandan.gangopadhyay@gmail.com, 
sunandan@bose.res.in}} \\
Department of Physics and Astrophysics,\\
West Bengal State University, Barasat, India\\
and\\
Visiting Associate in S.N.Bose National Centre \\for
Basic Sciences, Kolkata, India
}
\date{}

\maketitle

%%%%%%%%%%%%%%%%
\begin{abstract}
%%%%%%%%%%%%%%%%
\noindent We apply the covariant anomaly cancellation method
to compute the Hawking fluxes from the event and cosmic horizons 
of the Schwarzschild-de Sitter black hole. The derivation is new
from the existing ones as we split the space in three different regions 
(near to and away from the event and cosmic horizons) and write
down the covariant energy-momentum tensor using three step functions which
covers the whole region leading elegantly to the conditions
required to compute the Hawking fluxes from the event and cosmic horizons.
\\[0.3cm]
\\[0.3cm]
%{\bf PACS:} 11.10.Nx 

\end{abstract}
%%%%%%%%%%%%%%%%%%%%%%%%%%%%%%%%%%%%%%%%%%%%%%%%%%%%%%%%%%%%%%%

\noindent {\it{Introduction :}}

\noindent In the early seventies, Hawking \cite{Hawking:rv}, \cite{Hawking:sw} 
proposed that black holes evaporate
due to quantum particle creation and behave like thermal bodies
with an appropriate temperature. This is essentially 
a consequence of quantisation of matter in a background spacetime
having an event horizon. There are several derivations
of this effect \cite{gibbons}, \cite{fulling}, \cite{Parikh}, 
\cite{srinivasan}, \cite{parikh100}, \cite{medved}. 

\noindent Recently, Robinson, Wilczek and collaborators gave
an interesting method to compute the Hawking fluxes using
chiral gauge and gravitational anomalies \cite{rw}. 
The method was soon extended to the case of 
charged blackholes \cite{iso}. It rests on the fact that the effective
theory near the event horizon is a two dimensional chiral theory which,
therefore, has gauge and gravitational anomalies.
However, an unpleasant feature of \cite{rw}, \cite{iso}
was that whereas the expressions for chiral anomalies were
taken to be consistent, the boundary condition necessary to fix the
parameters were vanishing of covariant current and 
energy-momentum tensor at the event horizon. 
A more conceptually cleaner
and economical approach based on cancellation 
of covariant (gauge/gravitational) anomaly
has been discussed in \cite{rb}. Since the boundary condition involved
the vanishing of covariant current/energy-momentum tensor at the horizon,
it is more natural to make use of covariant expressions for
gauge and gravitational anomaly.
The generalization of this approach to non-Schwarzschild black holes
has been done in \cite{sunandan, sunandan1}.    

\noindent In this paper, we adopt the method in (\cite{rw}) to discuss
Hawking radiation from Schwarzschild-de Sitter blackhole. In \cite{jiang},
\cite{jiang1} this was done using consistent anomalies.
The novelty in this derivation apart from using 
covariant expressions for the energy-momentum tensor
throughout the analysis is that we split the space in three different regions 
(near to and away from the event and cosmic horizons) and write
down the covariant energy-momentum tensor using three step functions which
covers the whole region. This splitting using three step functions
is different from the earlier approaches \cite{jiang}, \cite{jiang1}.
The method elegantly lead to the conditions
required to compute the Hawking fluxes from the event and cosmic horizons.\\

% The discussion is specifically done for Hawking radiation from Garfinkle-Horowitz-
%Strominger (GHS) blackhole in string theory which is an example
%of the most general spherically symmetric blackhole spacetime 

\noindent {\it{Hawking radiation from Schwarzschild-de Sitter blackhole :}}\\

\noindent The Schwarzschild solution with a positive 
cosmological constant $\Lambda$ represents
a black hole in asymptotically de Sitter space. The metric of the four
dimensional Schwarzschild-de Sitter black hole reads
\begin{eqnarray}
ds^{2}=-f(r)dt^{2} + \frac{1}{f(r)}dr^{2} + r^{2}d\Omega
\label{2} 
\end{eqnarray}
where,
\begin{eqnarray}
f(r) &=& 1- \frac{2M}{r}-\frac{1}{3}\Lambda r^{2}\quad;\quad\Lambda>0~.
\label{2a}
\end{eqnarray}
It is easy to see that $f(r)=0$ at two positive values of $r$ when 
$9\Lambda M^{2}<1$. The smaller root $r_{H}$ denotes the position
of the event horizon and the larger one $r_{C}$ denotes the position
of the cosmic horizon.
With the aid of dimensional reduction near the event
and cosmic horizons, one can effectively describe
a theory with a metric given by the ``$r-t$" sector of the full 
spacetime metric (\ref{2}) near the two horizons. 

\noindent Now we divide the spacetime into three regions. 
In the region outside the event and cosmic horizons 
($r_{H}+\epsilon\leq r\leq r_{C}-\epsilon$),
the theory is free from anomaly and hence we have the 
energy-momentum tensor satisfying the conservation law
\begin{equation}
\nabla_{\mu}T^{\mu}_{(o)\nu} = 0~.\label{5}
\end{equation}
However, the omission of the ingoing modes in the 
region $r_{H}\leq r\leq r_{H}+\epsilon$ near the event horizon and the
omission of the outgoing modes in the 
region $r_{C}-\epsilon\leq r\leq r_{C}$ near the cosmic horizon leads to
an anomaly in the energy-momentum tensor in these regions. As we have
mentioned earlier, in this paper we shall focus only on the
covariant form of $d=2$ gravitational anomaly given by (\cite{rw, iso}):
\begin{eqnarray}
\nabla_{\mu}T^{\mu}_{(H)\nu} &=& \frac{1}{96\pi}
\epsilon_{\nu\mu}\partial^{\mu}R = \mathcal A_{\nu}\nonumber\\
\nabla_{\mu}T^{\mu}_{(C)\nu} &=& -\frac{1}{96\pi}
\epsilon_{\nu\mu}\partial^{\mu}R = -\mathcal A_{\nu}
\label{cov}
\end{eqnarray}
where, $\epsilon^{\mu\nu}$ and $\epsilon_{\mu\nu}$ are two
dimensional antisymmetric tensors for the upper and lower cases
with $\epsilon^{tr}=\epsilon_{rt}=1$.
It is easy to check that for the metric (\ref{2}), the 
anomaly is purely timelike with
\begin{eqnarray}
\mathcal A_{r} &=& 0\nonumber\\
\mathcal A_{t} &=& \partial_{r}N^{r}_{t}
\label{8}
\end{eqnarray}
where,
\begin{equation}
N^{r}_{t} = \frac{1}{96\pi}\left(ff''-\frac{f'^{2}}{2} \right).
\label{9}
\end{equation}
Now in the region free from anomaly, the conservation equation (\ref{5})
yields the differential equation
\begin{equation}
\partial_{r}T^{r}_{(o)t} = 0 \label{5a}
\end{equation}
which after integration leads to
\begin{equation}
T^{r}_{(o)t}(r) = a_{o}
\label{6}
\end{equation}
where, $a_{o}$ is an integration constant.
In the region near the event and cosmic horizons, the anomaly equations 
(\ref{cov}) lead to the following pair of differential equations
\begin{eqnarray}
\partial_{r}T^{r}_{(H)t}&=& \partial_{r}N^{r}_{t}(r)\nonumber\\
\partial_{r}T^{r}_{(C)t}&=& -\partial_{r}N^{r}_{t}(r)
\label{900}
\end{eqnarray}      
which after solution lead to
\begin{eqnarray}
T^{r}_{(H)t}(r) &=& b_{H} + N^{r}_{t}(r) 
- N^{r}_{t}(r_{H})\nonumber\\
T^{r}_{(C)t}(r) &=& c_{H} - N^{r}_{t}(r) + N^{r}_{t}(r_{H})
\label{10}
\end{eqnarray}
where, $b_{H}$ and $c_{H}$ are integration constants. 

\noindent Now as in (\cite{iso}, \cite{rb}),  
writing the energy-momentum tensor as a sum of three contributions
\begin{equation}
{T^{r}}_{t}(r) = T^{r}_{(H)t}(r)H(r)+T^{r}_{(o)t}(r)
[\theta(r-r_{H}-\epsilon)-\theta(r-r_{C}+\epsilon)]
+T^{r}_{(C)t}(r)\theta(r-r_{C}+\epsilon) 
\label{11}
\end{equation}
where, $ H(r) = 1 - \theta(r-r_{H}-\epsilon)$, we find 
\begin{eqnarray}
\nabla_{\mu}{T^{\mu}}_{t}
&=&\partial_{r}{T^{r}}_{t}(r)\nonumber\\
&=&\left(T^{r}_{(o)t}(r) - T^{r}_{(H)t}(r)
+N^{r}_{t}(r)\right)\delta(r-r_{H}-\epsilon) 
+\left(T^{r}_{(C)t}(r) - T^{r}_{(o)t}(r)
+N^{r}_{t}(r)\right)\delta(r-r_{C}+\epsilon)\nonumber\\
&&+\partial_{r}\left[N^{r}_{t}(r) H(r)\right]
-\partial_{r}\left[N^{r}_{t}(r) \theta(r-r_{C}+\epsilon)\right].
\label{12} 
\end{eqnarray}
The terms involving the total derivatives are cancelled by quantum effects of
classically irrelevant ingoing and outgoing modes at the event and
cosmic horizon respectively. The quantum effect to cancel these
terms is the Wess-Zumino term induced by the ingoing and outgoing 
modes near the event and cosmic horizons. Now since the full theory
must be invariant under diffeomorphism symmetry, hence the 
energy-momentum tensor must be covariantly conserved, i.e. 
$\nabla_{\mu}{T^{\mu}}_{t}=0$. This leads to two separate
conditions. When $r=r_{H}+\epsilon$, $\delta(r-r_{C}+\epsilon)$ 
vanishes and hence the coefficient of $\delta(r-r_{H}-\epsilon)$
must vanish at $r=r_{H}+\epsilon$ for 
$\nabla_{\mu}{T^{\mu}}_{t}$ to vanish leading to
\begin{eqnarray}
T^{r}_{(o)t}(r_{H}+\epsilon) - T^{r}_{(H)t}(r_{H}+\epsilon)
+N^{r}_{t}(r_{H}+\epsilon)=0~.
\label{12a} 
\end{eqnarray}
Similarly, when $r=r_{C}-\epsilon$, $\delta(r-r_{H}-\epsilon)$ 
vanishes and hence the coefficient of $\delta(r-r_{C}+\epsilon)$
must vanish at $r=r_{C}-\epsilon$ for 
$\nabla_{\mu}{T^{\mu}}_{t}$ to vanish leading to
\begin{eqnarray}
T^{r}_{(C)t}(r_{C}-\epsilon) - T^{r}_{(o)t}(r_{C}-\epsilon)
+N^{r}_{t}(r_{C}-\epsilon)=0~.
\label{12b} 
\end{eqnarray}
Now to compute the Hawking flux from the event horizon,
we focus our attention on (\ref{12a}) and also set the covariant boundary condition $T^{r}_{(H)t}(r_H)=0$
which yields $b_{H}=0$. Hence, eq.(\ref{12a}) becomes
\begin{eqnarray}
T^{r}_{(o)t}(r_{H}+\epsilon)=-N^{r}_{t}(r_H)
=\frac{1}{192\pi}f'^{2}(r_H)~.
\label{12aa} 
\end{eqnarray}
Now since $T^{r}_{(o)t}(r)=T^{r}_{(o)t}(r_{H}+\epsilon)=a_o$,
therefore the Hawking flux from the event horizon is given by
\begin{eqnarray}
a_o=\frac{1}{192\pi}f'^{2}(r_H)~.
\label{12aaa} 
\end{eqnarray}
To compute the Hawking flux from the cosmic horizon,
we focus our attention on (\ref{12b}) and also set the covariant boundary condition $T^{r}_{(C)t}(r_C)=0$
which yields $c_{H}=0$. Hence, eq.(\ref{12b}) becomes
\begin{eqnarray}
T^{r}_{(o)t}(r_{C}-\epsilon)=N^{r}_{t}(r_C)
=-\frac{1}{192\pi}f'^{2}(r_C)~.
\label{12bb} 
\end{eqnarray}
Now since $T^{r}_{(o)t}(r)=T^{r}_{(o)t}(r_{C}-\epsilon)=a_o$,
therefore the Hawking flux from the cosmic horizon is given by
\begin{eqnarray}
a_o=-\frac{1}{192\pi}f'^{2}(r_C)~.
\label{12bbb} 
\end{eqnarray}\\
%%%%%%%%%%%%%%%%%%%%%%%%%%%%%%%%%%%%%%%%%%%%%%
\noindent {\it{Discussions}} :\\

\noindent In this paper, we have computed the Hawking flux
from Schwarzschild-de Sitter blackhole which has two horizons
(event and cosmic) due to the presence of a 
positive cosmological constant. Unlike the approach in \cite{jiang},
we split the space in a different way so that the energy-momentum
tensor can be written down in terms of the
energy-momentum tensor near the event horizon (having a chiral anomaly), 
in the region away from the event and cosmic horizons (anomaly free
region) and near the cosmic horizon (having a chiral anomaly) 
using three step functions. Further, covariant expressions
for the energy-momentum tensor have been used throughout the paper.\\

\noindent {\it{Acknowledgements}} : The author would like
to thank the referee for useful comments.

%%%%%%%%%%%%%%%%%%%%%%%%%%%%%%%%%%%%%%%%%%%%%%%%%%%%%%%%%%%%%%%%%%


\begin{thebibliography}{99}
\bibitem{Hawking:rv}
S.~Hawking,
%``Black Hole Explosions,''
Nature (London) 248, 30 (1974).
\bibitem{Hawking:sw}
S.~Hawking,
%``Particle Creation By Black Holes,''
Commun.\ Math.\ Phys.\  43, 199 (1975).
\bibitem{gibbons}G. Gibbons, S. Hawking, Phys. Rev. D 15, 2752 (1977).
\bibitem{fulling}S.~Christensen, S.~Fulling,
Phys.\ Rev.\ D 15, 2088 (1977).
\bibitem{Parikh}
M.~Parikh and F.~Wilczek,
%``Hawking radiation as tunneling,''
Phys.\ Rev.\ Lett.\  85, 5042 (2000); [hep-th/9907001].
\bibitem{srinivasan}K.Srinivasan, T. Padmanabhan, Phys. Rev. D 60, 024007 (1999); [gr-qc/9812028].
\bibitem{parikh100}M. Parikh, Phys. Lett. B 546: 189, 2002;
[hep-th/0204107].
\bibitem{medved}A.M. Medved, Phys. Rev. D 66: 124009, 2002;
[hep-th/0207247].
\bibitem{rw} S. P. Robinson and F. Wilczek, Phys. Rev. Lett. 95,
 011303 (2005) [gr-qc/0502074].
\bibitem{iso} S. Iso, H. Umetsu and F.Wilczek, 
Phys. Rev. Lett. 96, 151302 (2006) [hep-th/0602146].
\bibitem{rb} R. Banerjee, S. Kulkarni, Phys. Rev. D 77: 024018, 2008; 
arXiv: 0707.2449, [hep-th].
\bibitem{sunandan}S. Gangopadhyay, S. Kulkarni, Phys. Rev. D 77: 
024038, 2008; arXiv: 0710.0974 [hep-th].
\bibitem{sunandan1}S. Gangopadhyay, Phys. Rev. D 78: 044026, 2008;
arXiv: 0803.3492 [hep-th].
\bibitem{jiang}Q.Q. Jiang, Class. Quantum Grav. 24 (2007) 4391; 
arXiv: 0705.2068 [hep-th]. 
\bibitem{jiang1}Q.Q. Jiang, S.Q. Wu, X. Cai, Phys. Lett. B 651: 65-70,2007;
arXiv: 0705.3871 [hep-th].


%\bibitem{tunneling}
%E.~C.~Vagenas,
%``Are extremal 2-D blackholes really frozen?,''
%Phys.\ Lett.\ B 503, 399 (2001).
%E.~C.~Vagenas,
%``Two-Dimensional Dilatonic Black Holes and Hawking Radiation,''
%Mod.\ Phys.\ Lett.\ A 17, 609 (2002);
%[arXiv:hep-th/0108147].
%%CITATION = HEP-TH 0108147;%%
%
%E.~C.~Vagenas,
%``Quantum corrections to the Bekenstein-Hawking entropy of the BTZ black
%hole via self-gravitation,''
%Phys.\ Lett.\ B 533, 302 (2002).
%\bibitem{bert} R. Bertlmann, ``Anomalies In Quantum Field Theory",
%(Oxford Sciences, Oxford, 2000). 
%\bibitem{bert1} R. Bertlmann and E. Kohlprath, Ann. Phys. (N.Y.) 
%288, 137 (2001).
%\bibitem{witt}L. Alvarez-Gaume, E. Witten, Nucl. Phys. B. 234, 269 (1984).
%\bibitem{fuji} K. Fujikawa and H. Suzuki, ``Path Integrals and Quantum
%Anomalies", (Oxford Sciences, Oxford, 2004).  


%\bibitem{Muratasoda1} K. Murata and J. Soda, Phys. Rev. D {\bf 74}, 
%044018 (2006), [hep-th/0606069].
%\bibitem{Vagenas} E. Vagenas and S. Das, JHEP {\bf 0610}, 025 (2006)
%[hep-th/0606077]. 
%\bibitem{Setare} M. R. Setare, Eur. Phys. J. C {\bf 49}, 865 (2007),
% [hep-th/0608080].
%\bibitem{chen} Z. Xu and B. Chen, Phys. Rev. D {\bf 75} 024041 (2007) 
%[hep-th/0612261].
%\bibitem{Morita} S. Iso, T. Morita and H. Umetsu, JHEP {\bf 0704},
% 068 (2007) [hep-th/0612286].
%\bibitem{Jiang} Q. Q. Jiang and S. Q. Wu, Phys. Lett. B 
%{\bf 647}, 200 (2007) [hep-th/0701002].
%\bibitem{Wu}Q. Q. Jiang, S. Q. Wu and X. Cai, [hep-th/0701048].
%\bibitem{Cai}Q. Q. Jiang, S. Q. Wu and X. Cai, Phys. Rev. D {\bf 75}, 064029 (2007) [hep-th/0701235].
%\bibitem{Zhang} X. Kui, W. Liu and H. Zhang, 
%Phys. Lett. B {\bf 647}, 482 (2007), [hep-th/0702199].
%\bibitem{Shin} H. Shin and W. Kim, arXiv:0705.0265 [hep-th]
%\bibitem{Peng} J. J. Peng and S. Q. Wu, arXiv:0705.1225 [hep-th].
%\bibitem{Jiang2} Q. Q. Jiang, arXiv:0705.2068 [hep-th].
%\bibitem{He} B. Chen and W. He, arxiv:0705.2984 [gr-qc].
%\bibitem{Murata2} U. Miyamoto and K. Murata, arXiv:0705.3150 [hep-th].
%\bibitem{Isomorita} S. Iso, T. Morita and H. Umetsu, arXiv: 0705.3494
% [hep-th].
%\bibitem{das1} S. Das, S.P Robinson and E. C Vagenas, arXiv: 0705.2233

%\bibitem{isorecent} S. Iso, T. Morita, H. Umetsu, 
%arXiv: 0710.0456, [hep-th].
%\bibitem{rb2} R. Banerjee, S. Kulkarni, e-print: arXiv: 0709.3916, [hep-th].
%\bibitem{gibb}G.W. Gibbons, Nucl. Phys. B. 207, 337 (1982);
%G.W. Gibbons, K. Maeda, Nucl. Phys. B. 298, 741 (1988).
%\bibitem{ghs}D. Garfinkle, G.T. Horowitz, A. Strominger, Phys. Rev. D. 43,
%3140 (1991) [Erratum-ibid. D. 45, 3888 (1992)].
%\bibitem{das}E. C. Vagenas, S. Das, JHEP 0610: 025, 2006, [hep-th/0606077].
%\bibitem{hms}G.~T.~Horowitz, J.~M.~Maldacena and A.~Strominger,
%Phys.\ Lett.\  B {\bf 383} (1996) 151, [arXiv:hep-th/9603109].
%\bibitem{mss}J.~Maharana and J.~H.~Schwarz,
%Nucl.\ Phys.\  B {\bf 390} (1993) 3, [arXiv:hep-th/9207016];
%A.~Sen, Nucl.\ Phys.\  B {\bf 404} (1993) 109, [arXiv:hep-th/9207053].
%\bibitem{D1}H. Shin, W. Kim, JHEP 0706 (2007), 012, [arXiv:hep-th/0705.0265].
%%%%%%%%%%%%%%%%%%%%%%%%%%%%%%%%%%%%%%%%%%%%%%%%%%%%%%%%%%%%%%%%%%%%%%%
 
\end{thebibliography}
\end{document}